\documentstyle[12pt,twoside,fleqn,espcrc1,epsfig]{article}

\def\bra {\langle} 
\def\ket {\rangle} 
\newcommand{\AmS}{{\protect\the\textfont2
  A\kern-.1667em\lower.5ex\hbox{M}\kern-.125emS}}

\title{Constituent quark model for baryons with strong 
quark-pair correlations and non-leptonic weak transitions of hyperon}

\author{K.~Suzuki\address{RCNP, Osaka University,  
        Osaka 567-0047, Japan}%
        \thanks{Talk presented at International Symposium on `Physics 
of Hadrons and Nuclei' held at Tokyo, Japan, Dec.~14-17, 1998
e-mail: ksuzuki@rcnp.osaka-u.ac.jp}, %
        E.~Hiyama\address{RIKEN, Wako, Saitama 351-0198, Japan}, 
        H.~Toki$^a$, 
	M.~Kamimura\address{Department of Physics, Kyushu University, 
        Fukuoka 812-8581, Japan}}

\begin{document}
% typeset front matter
\maketitle

\begin{abstract}
We study the roles of quark-pair correlations 
for baryon properties, in particular on non-leptonic weak decay 
of hyperons.  
We construct the quark wave function of baryons by solving the three 
body problem explicitly with confinement force and 
the short range attraction for a pair of quarks 
with their total spin being 0.  
We show that the existence of the strong quark-quark correlations enhances the 
non-leptonic 
transition amplitudes which is consistent with the data, 
while the baryon masses and radii are kept to the experiment.   

\end{abstract}

\section{Introduction}

Properties of light baryons have been extensively studied based on the 
constituent quark picture, in which 
the constituent quarks are assumed to be 
identified with quasi-particles of QCD vacuum.  Their results 
including applications for the two nucleon system are quite consistent 
with experiments.  
%although this simple model involves several defects and 
%adjustable assumptions. 
%In this picture, baryons consist of three valence quarks which 
%interact each other with the confinement force and short range 
%spin-dependent interaction.  
Despite the success of this approach, understandings of 
the non-leptonic weak hyperon 
decay and its $\Delta I = 1/2$ rule are still incomplete\cite{Review,model}.

It is well known that the factorization as well as the penguin 
contributions are too small to reproduce the $\Delta I=1/2$ decays of 
hyperons.  
Making use of the soft pion theorem, one can derive the 
baryon pole contribution 
to the non-leptonic decay, where initial hyperon $B_i$ changes to the 
intermediate state baryon $B_n$ by the weak interaction and then 
$B_n$ emits the pion to produce the final state $B_f$(and vice versa).  
The pole approximation with SU(6) spin-flavor symmetry for SU(3) baryons 
can reproduce relative magnitudes of various hyperon decays very 
well.  
However, if one calculates these transition amplitudes using the 
constituent quark model, the absolute value of the 
amplitudes is about a half of the experimental data at most\cite{model}.

Let us write the effective weak 
interaction Hamiltonian\cite{Hamiltonian};
\begin{equation}
{\cal H}_{W}={{G_F\sin \theta \cos \theta } \over {\sqrt 2}}
\sum\limits_i^{} {c_i(\mu ^2)\;}O_i + \mbox{h.c.}
\end{equation}
where
\begin{eqnarray}
O_1&=&[\bar u\gamma _\mu (1-\gamma _5)s][\bar d\gamma _\mu (1-\gamma _5)u] \\
O_2&=&[\bar d\gamma _\mu (1-\gamma _5)s][\bar u\gamma _\mu (1-\gamma _5)u] \; .
\end{eqnarray}
Here, we write dominant operators only.  Within the baryon pole 
approximation, the parity-conserving transition amplitude 
$B_i \to B_f + \pi ^a$ is given by, 
\begin{equation}
B_{fi}^a = \sum\limits_n {{{M_i+M_f} \over 
{f_\pi }}}\left( {{{G_{fn} ^a W_{ni}} \over {M_i-M_n}}+{{W_{fn} 
G_{ni}^a} 
\over {M_f-M_n}}} \right)\;\;,
\end{equation}
where matrix elements are defined as, 
\begin{eqnarray}
&&\left\langle {B_f|A^{a\mu }|B_n} \right\rangle =G_{fn}^a\bar 
u(f)\gamma ^\mu \gamma ^5u(n) \\
&&\left\langle {B_n|{\cal H}_W|B_i} \right\rangle =h_{ni}\bar u(n)u(i)
\end{eqnarray}
Axial-vector coupling constants $G_{fn}$ are rather well known quantities.  
Hence, we face to determine the expectation values of the weak 
Hamiltonian $h_{ni}$ by the quark model.  
We recall that the quark models such as Isgur-Karl HO model or 
MIT bag model give much smaller values than the experimental 
data\cite{model,Bando}.  
It is instructive to rewrite the operator (2,3) in the non-relativistic limit 
as
\begin{equation}
O_1, \, O_2 \to 
a_d^\dag \,  a_u^\dag (1 - \vec \sigma_u \cdot \vec \sigma_s)
a_u  a_s
\end{equation}
where $a_i$, $ a_i^\dag$ are annihilation and creation operators of 
quarks with 
flavor $i$.  Presence of the spin-projection operator 
$(1 - \vec \sigma_u \cdot \vec \sigma_s)$ tells us that only the $us$ pair 
with their total spin being 0 can contribute to the weak decay process,   
namely, the weak transition is generated by the two 
body process between spin-0 quark pairs; $(us)^0 \to (ud)^0$.  
%The isospin of the initial $(us)^0$ pair is $1/2$, and the final $(ud)^0$ 
%has the isospin-0 due to the antisymmetrization.  
%Thus, this process guarantees the $\Delta I=1/2$ dominance 
%in the non-leptonic 
%hyperon decays as pointed out long ago\cite{Review}.  
Now it is clear that this decay amplitude is very sensitive to the 
correlation of the spin-0 quark pair in the baryons\cite{Diquark}.  
The standard constituent quark model never incorporates such a 
correlation properly.  
However, in fact, more fundamental studies on non-perturbative QCD, 
e.g.~instanton liquid model\cite{Instanton}, suggest that there exists 
the strong attractive 
correlation for the quark-quark pair with $s=0$.  
These considerations naturally lead us to study the quark structure 
of baryons by taking into account the attractive correlation
which could enhance the weak decay amplitudes.

\section{Constituent quark model for baryons with spin dependent correlation}

Our purpose here is to construct the simplified quark model to deal with 
the quark pair correlations and thus account for the 
non-leptonic weak decay.  
We phenomenologically introduce the effective 
Hamiltonian which includes the confinement force and the spin-dependent 
part as; 
\begin{eqnarray}
{\cal H} &=&\sum\limits_i {{{p_i^2} \over {2m_i}}}+V_{C}+V_{S} + V_0 \\
V_{C} &=&\sum\limits_{i<j} {{1 \over 2}K\left( {\vec r_i- \vec 
r_j} \right)^2} \\
V_{S}&=&\left\{ {\matrix{{0\quad (s=1\;\mbox{pair})}\cr
{\;\sum\limits_{i<j} {{{C_{SS}} \over {m_i m_j}}}
\mbox{Exp}\left[ {-\left( {\vec r_i-\vec 
r_j} \right)^2  / \beta ^2} \right]}\cr
}} \right.\quad (s=0\;\mbox{pair})
\end{eqnarray}
where $m_i$ are the constituent quark masses,  and $K, C_{SS}, \beta$ 
are the model parameters.  $V_0$ is the constant parameter which 
contributes to the over all shift of the resulting spectrum and is chosen 
to adjust the energy of the lowest state to the nucleon mass.   
Constituent quark masses are taken to be 
$m_u = m_d = 330 \mbox{MeV}$ and $m_s = 500 \mbox{MeV}$.

Using this Hamiltonian, we shall solve non-relativistic three 
body problem rigorously.  
We use the coupled-rearrangement-channel 
variational method with infinitesimally-shifted-Gaussian-Lobe 
basis functions which is developed by ones of the authors\cite{Hiyama}. 
We assume only the isospin symmetry between up and down quarks.  
The quark wave functions are constructed by the antisymmetrization 
without any further approximations or assumptions.  
We note that the SU(6) spin-flavor symmetry 
should be broken within our 
formalism because of the spin-dependent correlation.  
It is interesting to clarify the effects of the broken SU(6) on 
the non-leptonic weak decay and other baryon properties.  

\vspace{-0.5cm}

\begin{figure}[htb]
\begin{minipage}[c]{100mm}
%\framebox[79mm]{\rule[-26mm]{0mm}{52mm}}
\psfig{file=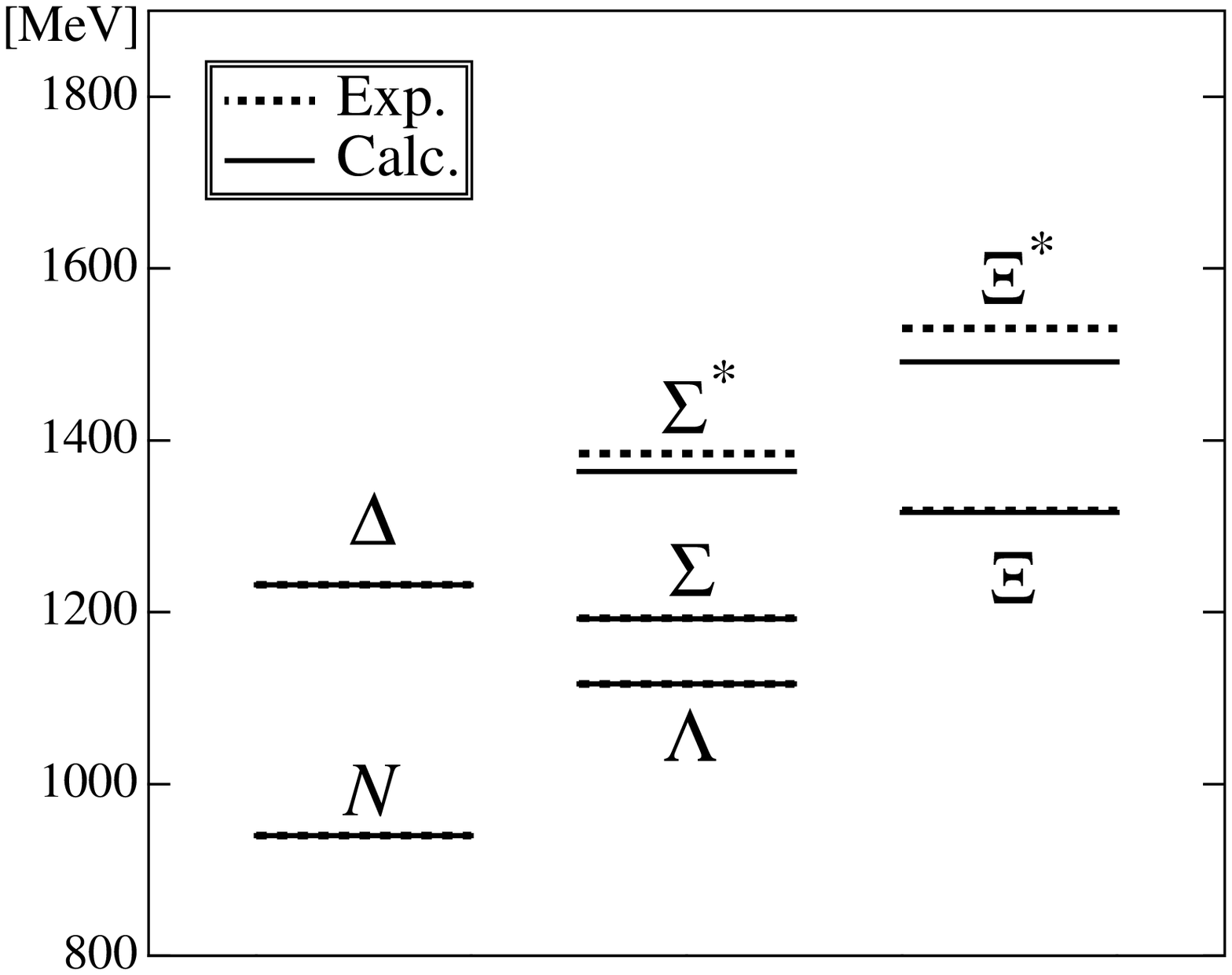,height=2.7in} 
\vspace{-0.8cm}
\caption{SU(3) baryon mass spectrum: Calculations are shown by the 
solid lines, and experiments by the dashed ones. }
\label{fig1}
\end{minipage}
\hspace{\fill}
\begin{minipage}[c]{53mm}
%\framebox[74mm]{\rule[-26mm]{0mm}{52mm}}
\psfig{file=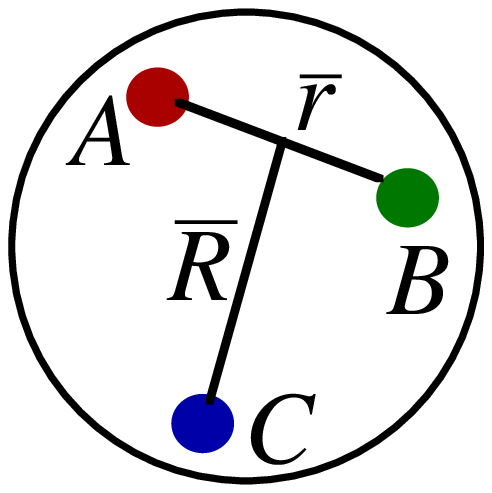,height=1.3in} 
\caption{Configuration of three quarks in nucleon}
\label{fig2}
\end{minipage}
\end{figure}

\vspace{-0.5cm}

\section{Results and discussions}

We shall fix the model parameters so as to reproduce the 
nucleon and $\Delta$ masses, proton radius and strength of the 
non-leptonic hyperon decay.  The experimentally measured proton radius 
includes contributions from both valence quark core part and its meson 
clouds.  
It is reasonable to subtract the vector meson dominance contribution 
$6 / m_\rho ^2$ from the 
experimental data of the proton electric radius $(0.86 \mbox{fm})^2$ 
to obtain the valence quark core radius $\bra r^2 \ket _{core}$.  
From this analysis, we determine $\bra r^2 \ket  _{core} =  
(0.6 \mbox{fm})^2$.  
We obtain $K=0.005 \mbox{GeV}^3$, $\beta = 
0.5 \mbox{fm}$ and $C_{SS} / m_u^2 = 1.4 \mbox{GeV}$.

We show first in Fig.1 SU(3) baryon mass spectrum.  
The agreement with the data encourages us to proceed our approach.  
In order to clarify the effects of the correlation on the nucleon 
structure, we calculate average distances of the 
Jacobi coordinate $\bar r$ and $\bar R$ 
defined in Fig.2.  We find $\bar r = 0.92 \mbox{fm}$ and 
$\bar R = 0.97 \mbox{fm}$
when the total spin $s$ of the quark pair $A$ and $B$ is zero, while 
$\bar r = 1.1 \mbox{fm}$ and $\bar R = 0.81 \mbox{fm}$ in the  $s=1$ case.  
Apparently, the quark correlation modifies quark distribution in the nucleon.  
%However, it seems that the correlation is not so strong to form 
%the so called `diquark'-clustering in the nucleon\cite{Diquark}.   
%Inclusion of the relativistic effects may change the present results, 
%and is the subject of future study.  

The weak matrix elements are shown in Table 1.  
In the left column we show the matrix elements with the quark 
correlation and the ones without the correlations in the right column.  
In the absence of the correlation $C_{SS} = 0$, 
a ratio $\bra p | H_W | \Sigma^+ \ket / 
\bra  n | H_W | \Lambda \ket = -2.45$ shows a perfect agreement with the SU(6) 
expectation $\sqrt{6} \simeq - 2.4494\cdots$.  
In the realistic case with the spin-dependent force, one can see the 
substantial 
enhancement of the matrix elements and the SU(6) breaking effects.  
The non-leptonic weak transition amplitudes are tabulated in table 2.  
We shown the pole contributions only in the second column and additional 
factorization and penguin contributions are in the third column.  Then, 
we show the total 
decay amplitudes in the forth column to be compared with the 
experiments.  
We find a good agreement for $\Sigma \to N \pi$ decays.  The pole contribution 
for $\Lambda \to n \pi^0$ is small and thus the total amplitude becomes 
about a half of the data.  This is because the SU(6) breaking effects on our 
quark wave function tends to reduce the matrix element 
$\bra n | H_W | \Lambda \ket$ relatively.  Improvement of this difficulty is 
now in progress.

\vspace{-1.cm}

\begin{table}[hbt]

% -----------------------------------------------------
% adapted from TeX book, p. 241
\newlength{\digitwidth} \settowidth{\digitwidth}{\rm 0}
\catcode`?=\active \def?{\kern\digitwidth}
% -----------------------------------------------------

\begin{center}

{\bf Table 1} Matrix elements of the weak Hamiltonian 

\vspace{0.2cm}

\begin{tabular*}{7cm}{@{}l@{\extracolsep{\fill}}|cc}
& with $V_{S}$ & without  $V_{S}$ \\
\hline
$\bra n | {\cal H}_W | \Lambda \ket $   & $-0.546$ & $-0.179$ \\
$\bra p | {\cal H}_W | \Sigma ^+ \ket $ & 1.65 & 0.440 \\
\end{tabular*}

\vspace{0.6cm}

{\bf Table 2} P-wave non-leptonic weak transition amplitude 

\vspace{0.2cm}

\begin{tabular*}{11cm}{@{}l@{\extracolsep{\fill}}|ccc|c}
Decay & Pole & others  & total & Exp. \\
\hline
$\Sigma \to p \pi^0$ & $ 26.0$   & $2.05$  & 28.05   & $26.24 \pm 1.32$ \\
$\Sigma^+ \to n \pi^+$ & $ 43.4$ & 0.0     & 43.3    & $41.83 \pm 0.17$ \\
$\Lambda \to n \pi^0$  & $-2.30$ & $-5.02$ & $-7.32$ & $ -15.61 \pm 1.40$
\end{tabular*}

%\end{minipage}

\end{center}

\end{table}

\vspace{-1cm}

In conclusion, we have studied the roles of the spin-dependent correlations 
in the baryons for the non-leptonic weak transitions of hyperons.  
We have constructed the simple non-relativistic quark model with the 
attractive correlation for the spin-0 quark pair, and  solved the three body 
problem rigorously.  
Our model reproduces the non-leptonic hyperon decays as well as the 
baryon masses and radius due to the existence of the spin-0 quark pair 
correlations.

\noindent
%{\bf Acknowledgments}
%
%Yazaki......

\end{document}